%% file: main.tex
\documentclass[sn-basic,iicol]{sn-jnl}
\usepackage[utf8]{inputenc}
\usepackage[english]{babel}
\usepackage{xcolor}
\usepackage{multirow}
\usepackage{ulem}
\newcommand{\bs}{\boldsymbol}

\DeclareMathOperator{\Imag}{Im}
\DeclareMathOperator{\Sigm}{Sigm}

\usepackage{etoolbox}
\makeatletter
\patchcmd{\ps@headings}
{\hbox to \hsize{\hfill Springer Nature 2021 \LaTeX\ template\hfill}}
{\hbox to \hsize{}}
{}
{}
\patchcmd{\ps@headings}
{\hbox to \hsize{\hfill Springer Nature 2021 \LaTeX\ template\hfill}}
{\hbox to \hsize{}}
{}
{}
\patchcmd{\ps@titlepage}
{\hbox to \hsize{\hfill Springer Nature 2021 \LaTeX\ template\hfill}}
{\hbox to \hsize{}}
{}
{}
\makeatother

\title[a]{Comparison between an exact and a heuristic neural mass model with second order synapses}
\author*[1]{\fnm{Pau} \sur{Clusella}}\email{pau.clusella@upf.edu}
\author[2]{\fnm{Elif} \sur{K\"oksal-Ers\"oz}}
\author[1]{\fnm{Jordi} \sur{Garcia-Ojalvo}}
\author*[3]{\fnm{Giulio} \sur{Ruffini}}\email{giulio.ruffini@neuroelectrics.com}
\affil[1]{\orgdiv{Department of Medicine and Life Sciences}, \orgname{Universitat Pompeu Fabra}, \orgaddress{\street{Barcelona Biomedical Research Park}, \city{Barcelona}, \postcode{08003}, \country{Spain}}}
\affil[2]{\orgdiv{LTSI - UMR 1099}, \orgname{INSERM, Univ Rennes}, \orgaddress{\street{Campus Beaulieu},  \city{Rennes}, \postcode{F-35000}, \country{France}}}
\affil[3]{\orgdiv{Brain Modeling Department}, \orgname{Neuroelectrics}, \orgaddress{\street{Av. Tibidabo, 47b  },  \city{Barcelona}, \postcode{08035}, \country{Spain}}}

\abstract{
Neural mass models (NMMs) are designed to reproduce the collective dynamics of neuronal populations.
A common framework for NMMs assumes heuristically that the output firing rate of a neural population can be described by a static nonlinear transfer function (NMM1).
However, a recent exact mean-field theory for quadratic integrate-and-fire (QIF) neurons challenges this view by showing that the mean firing rate is not a static function of the neuronal state but follows two coupled nonlinear differential equations (NMM2).
Here we analyze and compare these two descriptions in the presence of second-order synaptic dynamics. 
First, we derive the mathematical equivalence between the two models in the infinitely slow synapse limit, i.e., we show that NMM1 is an approximation of NMM2 in this regime.
Next, we evaluate the applicability of this limit in the context of realistic physiological parameter values by analyzing the dynamics of  models with inhibitory or excitatory synapses. We show that
NMM1 fails to reproduce important dynamical features of the exact model, such as the self-sustained oscillations of an inhibitory interneuron QIF network. Furthermore, in the exact model but not in the limit one, stimulation of a pyramidal cell population induces resonant oscillatory activity whose peak frequency and amplitude increase with the self-coupling gain and the external excitatory input.  
This may play a role in the enhanced response of densely connected networks to weak uniform inputs, such as the electric fields produced by non-invasive brain stimulation. 
}

\jyear{2022}%

\begin{document}

\maketitle
\clearpage

\input{introduction}

\input{basic_models}

\input{slowfast_synapses}

\input{dynamics}

\input{conclusions}

\bmhead{Acknowledgments}
 This work has received funding from the European Research Council (ERC Synergy) under the European Union’s Horizon 2020 research and innovation programme (grant agreement No 855109) and from the Future and Emerging Technologies Programme (FET) under the European Union’s Horizon 2020 research and innovation programme (grant agreement No 101017716).
 J.G.-O. is supported by the Spanish Ministry of Science and Innovation and FEDER (grant PGC2018-101251-B-I00), by the ``Maria de Maeztu'' Programme for Units of Excellence in R\&D (grant CEX2018-000792-M), and by the Generalitat de Catalunya (ICREA Academia programme).

\section*{Declarations}

{\bf Conflicts of interests:} GR is a co-founder of Neuroelectrics, a Company that manufactures tES and EEG technology. The remaining authors don’t have any competing interests.

\input{appendix}

\bibliography{references}

\end{document}

%% file: introduction.tex
\section{Introduction}

Neural mass models (NMMs) provide a physiologically grounded description of the average synaptic activity and firing rate of neural populations \citep{wilson_excitatory_1972,lopes_da_silva_model_1974,lopes_da_silva_model_1976,jansen_neurophysiologically-based_1993,jansen_electroencephalogram_1995,wendling_epileptic_2002}.
First developed in the 1970s, these models are increasingly used for both local and whole-brain modeling in, e.g., epilepsy \citep{wendling_epileptic_2002,wendling_transition_2008,jedynak2017temporally} or Alzheimer's disease \citep{pons2010relating,Stefanovski:2019aa}, and for understanding and optimizing the effects of transcranial electrical stimulation (tES) \citep{molaee-ardekani_computational_2010,merlet_oscillatory_2013,kunze_transcranial_2016,Ruffini:2018aa, sanchez-todo_personalization_2018}.
 However, they are  only partly derived from first principles.
 While the post-synaptic potential dynamics are inferred from data and can be grounded on diffusion physics \citep{Destexhe:1998aa,Pods:2013aa,Ermentrout:2010aa}, the transfer function linking the mean population membrane potential with the corresponding firing rate (Freeman's ``wave-to-pulse'' sigmoid function) rests on a weaker theoretical standing \citep{wilson_excitatory_1972,Freeman:1975aa,Kay:2018aa, Eeckman:1991aa}.
 This results in  a limited understanding on the range of applicability of the theory. For example, although models for the effects of an electric field at the single neuron are now available \citep{aberra_biophysically_2018, Galan:2021aa}, it is unclear how they should be used at the population-level representation.
 
In 2015, Montbri\'o, Paz\'o, and Roxin (MPR)~\citep{Montbrio:2015aa}  derived an exact mean-field theory for networks of quadratic integrate-and-fire (QIF) neurons, thereby connecting microscale neural mechanisms with mesoscopic brain activity.
 Within this framework, the response of a neural population is described by a low-dimensional system representing the dynamics of the firing rate and mean membrane potential.
Therefore, the MPR equations can be seen to replace the usual static transfer sigmoid function with two differential equations grounded on the biophysics of the single neurons.
Since then, the theory has been applied to cover increasingly complex formulations of the single-neuron activity, including
time-delays~\citep{pazo2016,devalle2018,ratas2018}
dynamic synapses~\citep{Montbrio:2015aa,ratas2016,Devalle:2017aa,Dumont:2019aa,coombes2019,byrne2020,byrne2022},
gap-junctions~\citep{laing2015,pietras2019},
stochastic fluctuations~\citep{ratas2019,goldobin2021,clusella2022},
asymmetric spikes~\citep{montbrio2020},
sparse connectivity~\citep{divolo2018,bi2021},
and short-term plasticity~\citep{taher2020,taher2022}. 

In the limit of very slow synapses, the firing rate of the MPR formulation can be cast as a static
function of the input currents, in the form of a population-wide $f$-$I$ curve \citep{Devalle:2017aa}. 
This function can be used to derive an NMM with exponentially decaying synapses, which fails to reproduce the dynamical
behavior of the exact mean-field theory, highlighting the importance of the dynamical equations in the MPR model \citep{Devalle:2017aa}.
In fact, empirical evidence suggests that post-synaptic currents  display rise and a decay time scales~\citep{lopes_da_silva_model_1974,Jang:2010aa}.
These type of synaptic dynamics can be modelled through a second-order linear equation, which forms the basis for many NMMs (see e.g. \cite{lopes_da_silva_model_1974,jansen_neurophysiologically-based_1993,wendling_epileptic_2002}).
This has been also noticed by other researchers, who have recently studied exact NMMs with second-order synapses~\citep{coombes2019,byrne2020,byrne2022}.
However, a formal comparison between the MPR formalism with second-order synaptic dynamics and classical, heuristic NMMs has not yet been established.

In this paper we analyze the NMM that results from applying the mean-field theory to a population of QIF neurons with second-order equations for the synaptic dynamics.
The resulting NMM, which we refer to as NMM2 in what follows, contains two relevant time scales: one for the post-synaptic activity and one for the membrane dynamics.
These two time scales naturally bridge the Freeman ``wave-to-pulse'' function with the nonlinear dynamics of the firing rate. 
In particular, following~\cite{Devalle:2017aa}, we show that, in the limit of very slow synapses and external inputs, the mean membrane potential and firing rate dynamics become nearly stationary.
This allows us to develop an analogous NMM with a static transfer function, which we will refer to as NMM1 for brevity.
Next, we analyze the dynamics of the two models using physiological parameter values for the time constants, in order to assess the validity of the formal mapping.
Bifurcation analysis of the two systems shows that the models are not equivalent, with NMM2 presenting a richer dynamical repertoire, including resonant responses to external stimulation in a population of pyramidal neurons, 
and self-sustained oscillatory states in inhibitory interneuron networks.

%% file: basic_models.tex
\section{Models}

\subsection{NMM with static transfer function}
Semi-empirical ``lumped" NMMs where first developed in the early 1970s by Wilson and Cowan~\citep{wilson_excitatory_1972}, Freeman~\citep{freeman1972,Freeman:1975aa}, and Lopes da Silva~\citep{lopes_da_silva_model_1974}.
This framework is based on two key conceptual elements.
The first one consists of the filtering effect of synaptic dynamics, which transforms the incoming activity (quantified by firing rate) into a mean membrane potential perturbation in the receiving population. 
The second element is a static transfer function that transduces the sum of the membrane perturbations from synapses and other sources into
an output mean firing rate (see \cite{grimbert_analysis_2006} for a nice introduction to the Jansen-Rit model).
We next describe these two elements separately.

The synaptic filter is instantiated by a second-order linear equation coupling the mean firing rate of arriving signals $r$ (in kHz) to the mean post-synaptic voltage perturbation $u$ (in mV)~\citep{grimbert_analysis_2006,Ermentrout:2010aa}:
\begin{equation}\label{eq:secondorder}
\tau_s^2 \ddot u = C\gamma r(t) -2\tau_s\dot u - u
\end{equation}
Here the parameter $\tau_s$ sets the delay time scale (ms), $\gamma$ characterizes the amplification factor in mV/kHz,
and $C$ is dimensionless and quantifies the average number of synapses per neuron in the receiving population. 
Upon inserting a single Dirac-delta-like pulse rate at time $t=0$, the solution of~\eqref{eq:secondorder} reads $u(t)=C\gamma \tau_s^{-2}te^{-t/\tau_s}$
for a system initially at rest ($\dot u(0)=u(0)=0$).
This model for PSPs activity is a commonly-used particular case of a more general formulation that
considers different rise and decay times for the post-synaptic activity~\citep{Ermentrout:2010aa}.

The synaptic transmission equation needs to be complemented by a relationship between the level of excitation of a neural population and its firing rate, namely, a transfer function, $\Phi$.
Through the transfer function, each neuron population converts the sum of its input currents, $I$,
to an output firing rate $r$ in a non-linear manner, i.e., $r(t) = \Phi[ I(t)]$.
Wilson and Cowan, and independently Freeman, proposed a sigmoid function as a simple model 
to capture the response of a neural mass to inputs, based on modeling insights and empirical observations~\citep{wilson_excitatory_1972,Freeman:1975aa, Eeckman:1991aa}.
A common form for the sigmoid function is
\begin{equation} \label{eq:sigmoid}
  \Sigm[ I] =  \frac{2e_0}{1+e^{\rho (I_{0}- I)}}\,,  
\end{equation}
where $e_0$ is the half-maximum firing rate of the neuronal population, $I_0$ is the threshold value of the input (when the firing rate is $e_0$), and $\rho$ determines the slope of the sigmoid at that threshold.
Beyond this sigmoid, transfer functions can be derived from specific neural models such as the leaky integrate-and-fire or the exponential integrate-and-fire, either analytically or numerically fitting simulation data, see e.g.~\cite{fourcaud-trocme2003,brunel2008,pereira2018,ostojic2011,carlu2020}.
In some studies, $\Phi$ is regarded as a function of mean membrane potential instead
of the input current~\citep{jansen_electroencephalogram_1995,wendling_epileptic_2002}. 
Nonetheless, the relation between input current and mean voltage perturbation
is often assumed to be linear, see for instance~\cite{Ermentrout:2010aa}.
Therefore, the difference between both formulations might be relevant only 
in the case where the transfer function has been experimentally or numerically derived.

The form of the total input current in Eq.~(\ref{eq:sigmoid}) will depend on the specific neuronal populations being considered, and on the interactions between them.
In what follows we focus on a single population with recurrent feedback and external stimulation. Hence the total input current is given as the contribution of three independent sources,
\begin{equation}\label{eq:arguments}
 I(t) = \kappa u(t) + p + I_E(t)
\end{equation}
where $\kappa$ is the recurrent conductance, $p$ is a constant baseline input current, and $I_E$ stands for the effect of an electric field.
Note that some previous studies do not use an explicit self-connectivity as an argument of the transfer function
(see e.g.~\cite{grimbert_analysis_2006,wendling_epileptic_2002,LopezSola2021}).
In the next section we show that the term $\kappa u(t)$ in~\eqref{eq:arguments} arises naturally
in recurrent networks.

Finally, 
we rescale the postsynaptic voltage by defining $s=u/(C\gamma)$,
and use the auxiliary variable $z$ to write Eq.~\eqref{eq:secondorder} as a system of two first-order differential equations.
With those choices, the final closed formulation for the neural population dynamics reads
\begin{equation}\label{eq:nmm1}
\begin{aligned}
        \tau_s \dot s &= z \\
        \tau_s \dot z &=  \Phi[ K s(t) + p + I_E(t)] - 2z -s \;
\end{aligned}
\end{equation}
where $K=C\gamma\kappa$.
We refer to this model in what follows as NMM1.


\subsection{Quadratic integrate-and-fire neurons and NMM2}
Consider a population of fully and uniformly connected QIF neurons indexed by $j=1, ..., N$.
The membrane potential dynamics of a single neuron in the population, $U_j$, is described by~\citep{latham2000,Devalle:2017aa}
\begin{equation} \label{eq:qif_phys}
c\dot U_j= g_L \frac{(U_j-U_r)(U_j-U_t)}{U_t-U_r} + I_{j,\text{total}}(t)\;,
\end{equation}
with $U_j$ being reset to $U_\text{reset}$ when $U_j \ge U_\text{apex}$.
In this equation, $U_r$ and $U_t>U_r$ represent the resting and threshold potentials of the neuron (mV),  $I_{j,\text{total}}$ the input current ($\mu$A),  $c$  the membrane capacitance ($\mu$F), and $g_L$ is the leak conductance (mS).  
If unperturbed, the neuron membrane potential tends to the resting state value $U_r$.
In the presence of input current, the membrane potential of the neuron $U_j$ can grow and surpass the threshold potential $U_t$, at which point the neuron emits a spike.
An action potential is produced when $U_j$ reaches a certain apex value $U_\text{apex}>U_t$,
at which point $U_j$ is reset to $U_\text{reset}$.

The total input current of neuron $j$ is
\begin{equation}
        I_{j,\text{total}}(t)=\chi_j(t) 
        + \kappa u(t) + \tilde I_E(t)\;.
\end{equation}
 The first term in this expression, $\chi_j(t)$, corresponds to a Cauchy white noise with median $\overline \chi$ and half-width at half-maximum $\Gamma$ (see~\cite{clusella2022}).
 The second term, $\kappa u(t)$, represents the mean synaptic transmission from other neurons $u(t)$, with coupling strength $\kappa$.
 As in NMM1, we assume that $u(t)$ follows Eq.~\eqref{eq:secondorder}.
 However, in this case, the firing rate is determined self-consistently from the population dynamics as
 \begin{equation}\label{eq:firingrate}
        r(t)=\frac{1}{N}\lim_{\tau_r\to 0}\sum_{j=1}^N \frac{1}{\tau_r}\sum_{k} \int_{t-\tau_r}^t \delta(t'-t_j^{(k)})\, d t'
\end{equation}
where $t_j^{(k)}$ is the time of the $k$th spike of neuron $j$, and the spike
duration time $\tau_r$ needs to assume small finite values in numerical simulations.
Finally, $\tilde  I_E(t)$ can represent both a common external current from other neural populations, or the effect of an electric field.
In the case of an electric field, the current can be approximated by $\tilde  I_E = \tilde  P\cdot E$, where $ \tilde  P$ is the dipole conductance term in the spherical harmonic expansion of the response of the neuron to an external, uniform electric field~\citep{Galan:2021aa}.
This is a good approximation if the neuron is in a subthreshold, linear regime and the field is weak, and can be computed using realistic neuron compartment models. 
We assume here for simplicity that all the QIF neurons in the population  are equally oriented with respect to the electric field (this could be generalized to a statistical dipole distribution).

In order to analyze the dynamics of the model it is convenient to cast it in a reduced form.
Following~\cite{Devalle:2017aa}, we define the new variables
\begin{equation}
        V_j=\left( U_j-\frac{U_r+U_t}{2}\right)/(U_t-U_r)\;, \:\:  s=\frac{u}{C\gamma}
\end{equation}
and redefine the system parameters (all dimensionless except for $\tau_m$) as
\begin{equation}
\begin{aligned}
        \tau_m&={c/g_L}\;,\\
        J&=\kappa \frac{C\gamma}{c(U_t-U_r)} \;,\\
        \eta&=\frac{\zeta}{g_L(U_t-U_r)}-\frac{1}{4}\;, \\
        \Delta&=\frac{\Gamma}{g_L(U_t-U_r)}\;, \\
        I_E(t)&=\frac{\tilde I_E(t)}{g_L(U_t-U_r)}\;,\text{ and} \\
        \xi_j(t)&=\frac{ \chi_j(t)}{g_L(U_t-U_r)}\;.
\end{aligned}
\end{equation}
With these transformations, the QIF model can be written as
\begin{equation}\label{eq:qif}
                \tau_m \dot V_j = V_j^2+\eta + J\tau_m s + \xi_j  + I_E(t) 
\end{equation}
with the synaptic dynamics given by
\begin{equation}\label{eq:synaptic}
\begin{aligned}
        \tau_s \dot s &= z \\
        \tau_s \dot z &=  r - 2z -s \;.
\end{aligned}
\end{equation}
These transformations express the QIF variables and parameters with respect to reference values of time ($c/g_L$), voltage ($U_t-U_r$), and  current ($g_L(U_t-U_r)$). In the new formulation, the only dimensional quantities have units of time ($\tau_m$ and $\tau_s$, in ms)
or frequency ($r$ and $s$, in kHz).
It is important to keep in mind these changes
when dealing with multiple interacting populations involving different parameters, and also when using empirical
measurements to determine specific parameter values.

\subsubsection{Exact mean-field equations with second order synapses (NMM2)}

Starting from Eq.~(\ref{eq:qif}),  \cite{Montbrio:2015aa}
 derived an effective theory of fully connected QIF neurons in the large $N$ limit.
Initially, the theory was restricted to deterministic neurons with Lorentzian distributed currents.
Recently it has also been shown to apply to neurons under the influence of Cauchy white noise,
a type of Lévy process that renders the problem analytically tractable~\cite{clusella2022}.
In any case, the macroscopic activity of a population of neurons given by Eq.~\eqref{eq:qif} can
be characterized by the probability of finding a neuron with membrane potential $V$ at time $t$, $P(V,t)$.
In the limit of infinite number of neurons ($N\to\infty$),
the time evolution of such probability density is given by a Generalized Master Equation (GME).
Assuming that the reset and threshold potentials for single neurons are set to $V_\text{apex}=-V_\text{reset}=\infty$,
the GME can be solved by considering that $P$
has a Lorentzian shape in terms of a time-depending mean membrane potential $v(t)$ and mean firing rate $r(t)$,
\begin{equation}
        P(V,t)=\frac{\tau_m r(t)}{[V-v(t)]^2+(\pi r(t) \tau_m)^2}\;,
\end{equation}
with
\begin{equation}
        \begin{aligned}\label{eq:nmm2}
                \tau_m \dot r &= \frac{\Delta}{\pi \tau_m} + 2rv\\
                \tau_m \dot v &= \eta - (\pi r \tau_m)^2 + v^2 +\tau_m J s + I_E(t)\;.\\
\end{aligned}
\end{equation}
Together with the synaptic dynamics~\eqref{eq:synaptic}, these equations describe
an exact NMM, which we refer to as NMM2.

%% file: slowfast_synapses.tex
\section{Slow and fast synapse dynamics limits}\label{section:slowfast}

\subsection{Slow synapse limit and map to NMM1}

Comparing the formulations of the semi-empirical model NMM1~(\ref{eq:nmm1}) and the exact mean-field
model NMM2 \eqref{eq:nmm2} one readily observes that the latter can be interpreted as an extension
of the former. The synaptic dynamics are given by the same equations in both models, yet in NMM2
the firing rate $r$ is not a static function of the input currents, but a system variable.
Moreover, NMM2 includes the dynamical effect of the mean membrane potential, $v$, which in the classical framework is assumed to be directly related with the post-synaptic potential.

\cite{Devalle:2017aa} showed that in a model with exponentially decaying (i.e. first-order) synapses, the
firing rate can be expressed as a transfer function in the limit of slowly decaying synapses.
Their work follows from previous results showing that, in class 1 neurons, the slow synaptic limit
allows one to derive firing rate equations for the population dynamics~\citep{ermentrout1994}.
Here we revisit the same steps to show that NMM2 can be formally mapped to a NMM1 form.
We perform such derivation in the absence of external inputs ($I_E(t)=0$).

Let us rescale time in Eq.~\eqref{eq:nmm2} to units of $\tau_s$, and the rate variables to units of $1/\tau_m$
using
\begin{equation}\label{eq:slowtransf}
 \tilde t=\frac{t}{\tau_s},\quad\tilde{s}=\tau_m s,\quad \tilde{z}=\tau_m z,\quad \tilde{r}=\tau_m r\;.
\end{equation}
Additionally we define $\epsilon=\tau_m/\tau_s$. Then the NMM2 model (Eqs.~\eqref{eq:synaptic} and~\eqref{eq:nmm2}) reads
\begin{equation}\label{eq:nmm2_eps}
        \begin{aligned}
                \epsilon \frac{d\tilde r}{d\tilde t} &= \frac{\Delta}{\pi} + 2\tilde rv\\
                \epsilon \frac{dv}{d\tilde t} &= \eta + v^2-(\pi \tilde r)^2 + J \tilde s \\
                \frac{d\tilde s}{d\tilde t} &= \tilde z \\
                \frac{d\tilde z}{d\tilde t} &= \tilde r - 2\tilde z -\tilde s \;.
\end{aligned}
\end{equation}
Taking now $\epsilon\to 0$ ($\tau_s \to \infty$), the equations for $\tilde r$ and $v$ become quasi-stationary in the slow time scale, i.e.
\begin{equation}
        \begin{aligned}
   0 &\approx \frac{\Delta}{\pi} + 2\tilde rv\\
   0 &\approx \eta + v^2-(\pi \tilde r)^2 +J \tilde s\;.
        \end{aligned}
\end{equation}
The solution of these equations is given by
$\tilde r=\Psi_\Delta\left( \eta+J \tilde s\right)$, where
\begin{equation}\label{eq:transfer}
     \Psi_\Delta(I)=\frac{1}{\pi\sqrt{2}}\sqrt{I+\sqrt{I^2 + \Delta^2}}\;.
\end{equation}
This is the transfer function of the QIF model, which relates input currents to the output firing rate.
Thus, in the limit $\epsilon\to 0$ system~\eqref{eq:nmm2_eps} formally reduces to the NMM1 formulation, Eq.~\eqref{eq:nmm1}.

In following sections we study to what extent this equivalence remains valid for finite ratios of $\tau_m/\tau_s$.
To that end, it is convenient to recast
the analogy between NMM1 and NMM2 in terms of the non-rescaled quantities, which corresponds to using
\begin{equation}\label{eq:equiv}
    \Phi = \tau_m^{-1} \Psi_\Delta,\; K= J\tau_m,\text{ and }p=\eta,
\end{equation}
in Eq.~\eqref{eq:nmm1}.

\begin{figure}[t!]
  \centerline{\includegraphics[width=0.5\textwidth]{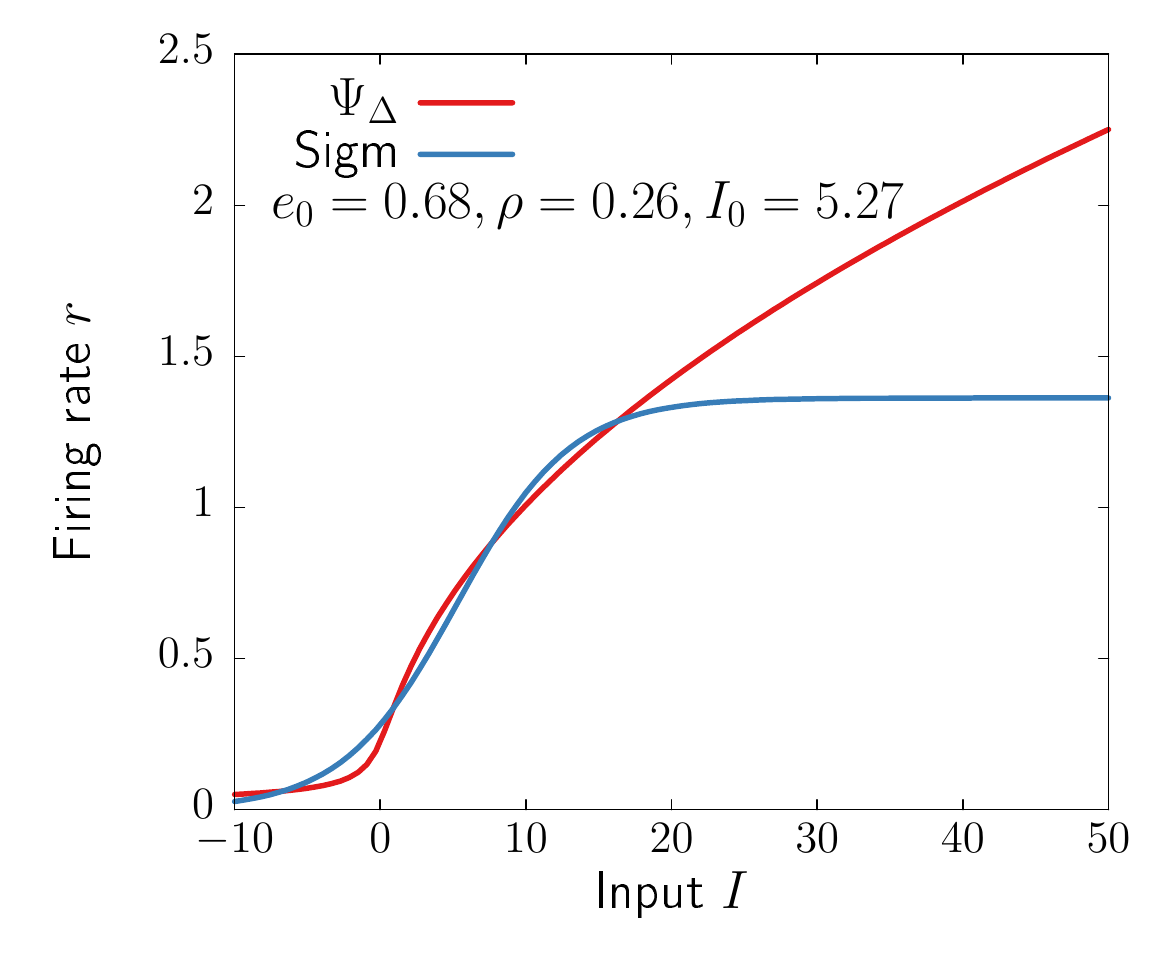}}
        \caption{  Transfer function of the QIF network~\eqref{eq:transfer} with $\Delta=1$ and the sigmoid~\eqref{eq:sigmoid} with parameters fitted to $\Psi_\Delta$.
        }
  \label{figure0}
\end{figure}

In Fig~\ref{figure0} we fit the parameters of the sigmoid function to $\Psi_\Delta$ for $\Delta=1$.
Despite the sudden sharp increase of both functions, there is an important qualitative difference:
the $f$-$I$ curve of the QIF model does not saturate for $I\to\infty$.
Other transfer functions derived from neural models share a similar non-bounded behavior~\citep{fourcaud-trocme2003,carlu2020}.
This reflects the continued increase of firing activity with increase input, which has been reported in experimental studies~\citep{Rauch2003:aa}.

\subsection{Fast synapse limit}

To explore the fast synapse limit it is convenient to rescale time as $\overline t=t/\tau_m$ in the NMM2 equations.
In this new frame, and defining $\delta:=\tau_s/\tau_m = 1/\epsilon$, the system reads
\begin{equation}\label{eq:nmm2_delta}
        \begin{aligned}
                \frac{d\tilde r}{d\overline t} &= \frac{\Delta}{\pi} + 2\tilde rv\\
                \frac{dv}{d\overline t} &= \eta + v^2-(\pi \tilde r)^2 + J \tilde{s} \\
                \delta \frac{d\tilde s}{d\overline t} &= \tilde z \\
                \delta \frac{d\tilde z}{d\overline t} &= \tilde r - 2\tilde z -\tilde s \;.
\end{aligned}
\end{equation}
where $\tilde r$, $\tilde s$ and $\tilde z$ are the rescaled variables defined in~\eqref{eq:slowtransf}.
With the algebraic conditions in the fast synapse limit, $\delta \to 0$ ($\tau_s \to 0$), Eq.~\eqref{eq:nmm2_delta} is reduced to 
\begin{equation}
        \begin{aligned}\label{eq:mpr}
                \frac{d\tilde r}{d\overline t} &= \frac{\Delta}{\pi} + 2\tilde rv\\
                \frac{dv}{d\overline t} &= \eta + v^2-(\pi \tilde r)^2 + J \tilde r \;,
\end{aligned}
\end{equation}
where we have used that $\tilde s=\tilde r$ as given by the synaptic equations.
 This is the model with instantaneous synapses analyzed by \cite{Montbrio:2015aa}, who showed that the $\eta$--$J$ phase diagram has three qualitatively distinct regions in the presence of a constant input: a single stable node corresponding to a low-activity state, a single stable focus (spiral) generally corresponding to a high-activity state, and a region of
 bistability between a low activity steady state and a regime of asynchronous persistent firing.

%% file: dynamics.tex
\section{System dynamics}\label{section:dynamics}

In the previous section we have shown that NMM2 can be mapped to NMM1 in the limit of slow synapses ($\tau_m/\tau_s\to 0$), using the scaling relations~\eqref{eq:equiv}. 
However, physiological values for the time constants might not be consistent with this limit.
Table~\ref{tab:taus} shows reference values for $\tau_m$ and $\tau_s$ corresponding to different neuron types and their corresponding neurotransmitters obtained from
experimental studies.
Notice that, in practice, such values also depend on the electrical and morphological properties of the neurons, and pre- and post-neuron types.
Such level of detail requires the use of conductance-based compartmental models,
a further step in mathematical complexity that is out of the scope of this paper.
Therefore, we take the values in Table~\ref{tab:taus} as coarse-grained quantities
that properly reflect the time scales in point neuron models such as the QIF~\eqref{eq:qif}
(for a more detailed discussion see Section~\ref{section:conclusions}).
In order to study to what extend these non-vanishing values of $\tau_m/\tau_s$ break down the equivalence
between the two models, in this section
we analyze and compare the dynamics of a single neural population with recurrent connectivity
described by both NMM1~(Eq.~\eqref{eq:nmm1} with Eq.~\eqref{eq:equiv}) and NMM2~(Eqs.~\eqref{eq:synaptic} and \eqref{eq:nmm2}).




\begin{table*}[ht]
    \caption{Values for the membrane time constants $\tau_m$
    and postsynaptic currents $\tau_s$, for
    Pyramidal neurons (Pyr), parvalbumin-positive (PV+),
    and neurogliaform cells (NGFC) \citep{neske2015,Zaitsev2012, Povysheva2007, Avermann2012, Olah2007, seay2020, karnani2016, bacci2003, Deleuze2019}.
    Notice that, in general, the synaptic time-constant should depend on the neurotransmitter, and
    the pre- and post-synaptic cells.
    Since we only consider self-coupled populations, we do not specify
    time-constants for transmission across populations of different types. 
    }
    \centering
    \begin{tabular}{|c|c|c|c|c|}\hline
    Neuron type & Neurotransmitter & $\tau_m$ (ms) & $\tau_s$ (ms) & $\tau_m/\tau_s$ \\\hline
     Pyr & Glutamate & 15  & 10 & 1.5\\
     PV+ & GABA  & 7.5 & 2  & 3.75 \\
     NGFC & GABA & 11  & 20 & 0.55\\ \hline
    \end{tabular}
    \label{tab:taus}
\end{table*}

The first step is to identify the steady states of the system.
Since we derived the transfer function~\eqref{eq:transfer} by assuming the $r$ and $v$ variables
of NMM2 to be nearly stationary, the fixed points of both models coincide and are given by
\begin{equation}\label{eq:fixedpoints}
\begin{aligned}
        \tau_m r_0&=\Psi_\Delta(\eta+\tau_m Jr_0),\\
        v_0 &=-\frac{\Delta}{2\tau_m \pi r_0},\\
        s_0 &= r_0,\;\\ 
        z_0 &=0.
\end{aligned}
\end{equation}
Moreover, $r_0$ and $v_0$ are the equilibrium points of the two-dimensional system analyzed by~\cite{Montbrio:2015aa}.
Notice that the only relevant parameters for the determination of the fixed points are $J$, $\eta$, and $\Delta$.
The time constant $\tau_m$ only acts as a multiplicative factor of $r_0$ (and $s_0$), and $\tau_s$ does not enter into the expressions of the steady states.

Even though the steady states of the three models (NMM1, NMM2, and the original system of \cite{Montbrio:2015aa}) are the same, their stability properties might be different, as we now attempt to elucidate. The eigenvalues controlling the stability of the fixed points in NMM1 are
\begin{equation}\label{eq:eigvals}
\lambda_\pm=\tau_s^{-1}\left(-1\pm\sqrt{J\Psi_\Delta'[\eta + J\tau_m s_0 ]}\right)
\end{equation}
A similar closed expression for NMM2 is complicated to obtain and, in any case,
there are no explicit expressions for the steady states.
Thus, we use in what follows the numerical continuation software AUTO-07p~\citep{AUTO} to obtain the corresponding bifurcation diagrams.
We analyze separately the dynamics of excitatory ($J>0$) and inhibitory ($J<0$) neuron populations in the two NMM models.

\subsection{Pyramidal neurons}\label{section:PN}

\begin{figure*}[ht]
  \centerline{\includegraphics[width=\textwidth]{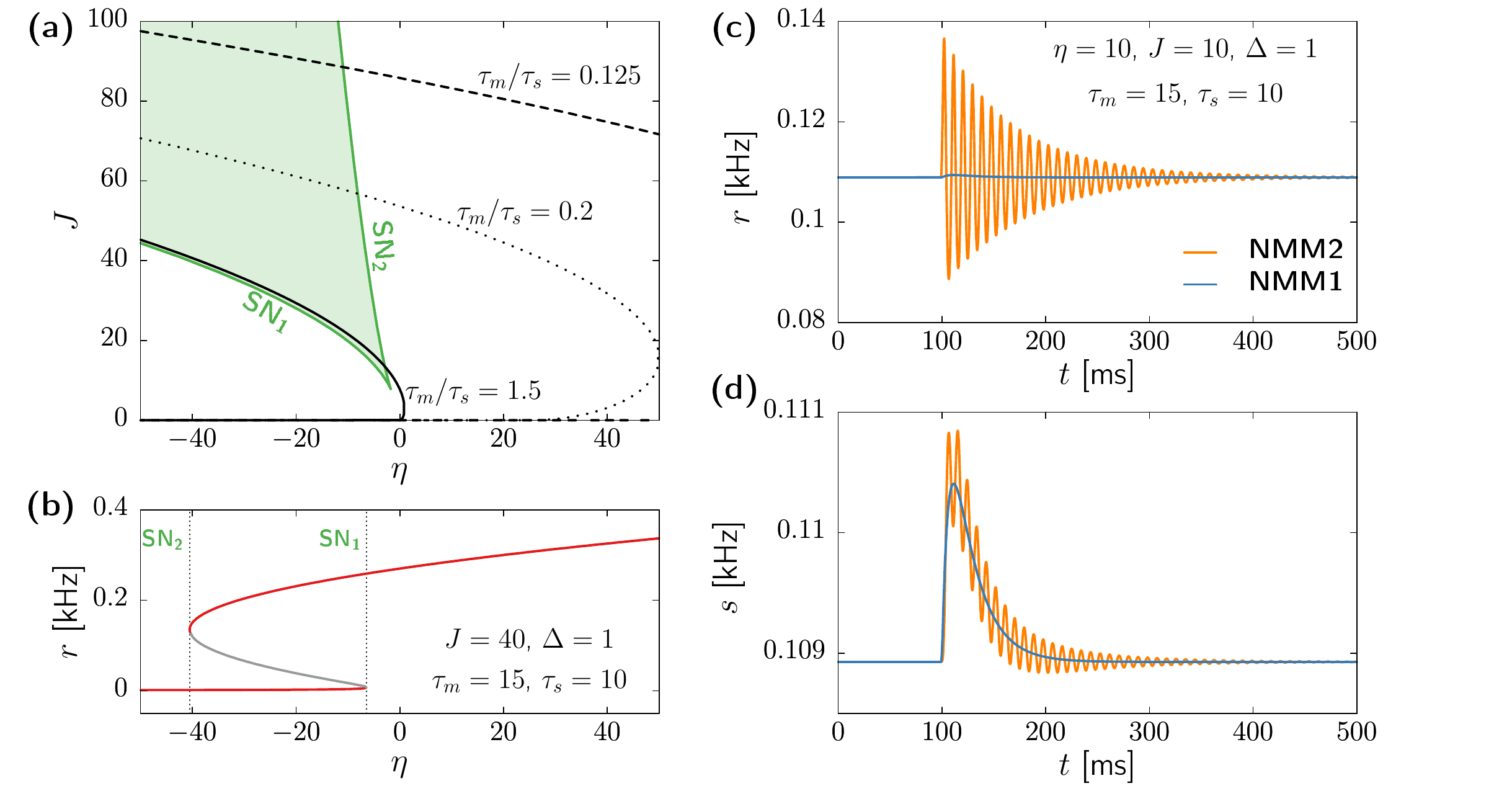}}
        \caption{
      {\bf Dynamics of a population of pyramidal neurons described by NMM1 and NMM2}.
        (a) Saddle-node bifurcations SN$_1$ and SN$_2$ (green curves) limiting the region of bistability (light-green region),
        and node-focus boundary for three different values of $\tau_m/\tau_s$ in NMM2 (solid, dotted and dashed black lines).
        (b) Steady-state value of the firing rate as a function $\eta$, for fixed $J=40$, $\Delta=1$, $\tau_m=15$~ms, and $\tau_s=10$~ms. The stable steady state branches are colored in red, and the unstable steady state branch in grey. Dashed vertical lines indicate the SN1 and SN2 bifurcation points (cf. panel a).
        (c,d) Time evolution of the firing rate $r$ (c) and synaptic variable $s$ (d) for NMM1 (blue) and NMM2 (orange), for $\eta=10$, $J=10$, $\Delta=1$, $\tau_m=15$ms, and $\tau_s=10$~ms. Initial conditions are at steady state, and a 1~ms-long pulse of $I_E(t)=10$ is applied at $t=100$~ms.
        }
  \label{fig:PN}
\end{figure*}

We start by analysing the dynamics of NMM1 in the case of excitatory coupling ($J>0$), by fixing $\Delta=1$ and varying $\eta$ and $J$.
Following Table~\ref{tab:taus}, we set $\tau_m=15$~ms and $\tau_s=10$~ms.
Since the NMM1 eigenvalues~\eqref{eq:eigvals}
are real for $J>0$, the fixed points do not display resonant behavior, i.e., they are either stable or unstable nodes.
For positive baseline input $\eta$, only a single fixed point exists irrespective of the value of the coupling $J$.
In contrast, a large region of bistability bounded by two saddle-node (SN) bifurcations emerges for negative $\eta$.
The green curves in Fig.~\ref{fig:PN}(a) show the two SN bifurcations, which merge in a cusp close to the origin of parameter space.
Within the region bounded by the curves (green shaded region), a low-activity and a high-activity state coexist, separated by a third unstable fixed point.
Figure~\ref{fig:PN}(b) displays, for instance, the stationary firing rate as a function of $\eta$ for $J=40$.
The values of the time constants $\tau_m$ and $\tau_s$ do not affect the bistability region.
However, the noise amplitude $\Delta$ does have an effect: as shown in Appendix~\ref{secA1}, NMM1
admits a parameter reduction that expresses all parameters and variables as functions of $\Delta$. 
Accordingly, the effects of modifying $\Delta$ on the stability of the fixed points are analogous to rescaling $\eta\to \eta/\Delta$ and $J\to J/\sqrt{\Delta}$ (see also \cite{Montbrio:2015aa}).
Therefore, the bistable region shrinks in the $(\eta,J)$ parameter space as the noise amplitude increases.

Since the fixed points of NMM1 and NMM2 coincide, these two branches of SN bifurcations also exist in NMM2.
Moreover, no other bifurcations arise, thus the diagrams depicted in Figs.~\ref{fig:PN}(a,b)
also hold for the exact model.
However, there is an important difference regarding the relaxation dynamics towards the fixed points:
While in NMM1 the steady states are always nodes, in NMM2 trajectories near the high-activity state might display transient oscillatory behavior.
Figures~\ref{fig:PN}(c,d) display, for instance, time series obtained from simulations of NMM1 (blue) and NMM2 (orange) starting at the fixed point, and receiving a small pulse applied at $t=100$~ms. 
Not only NMM2 displays an oscillatory response, but also the effect of the perturbation
in the firing rate is much larger in NMM2 than in NMM1.

Such resonant behavior of NMM2 corresponds to the two dominant eigenvalues
of the high-activity fixed point (those with largest real part) being complex conjugates of each other.
The black curves in Fig~\ref{fig:PN}(a) show the boundary line at which those two eigenvalues change from real (below the curves) to complex (above the curves).
For physiological values of $\tau_m$ and $\tau_s$ (continuous black line) this node-focus line remains very similar
to that of the model with instantaneous synapses studied in~\cite{Montbrio:2015aa}.
Reducing the ratio $\tau_m/\tau_s$ changes this situation.
As shown by the dotted and dashed curves in Fig.~\ref{fig:PN}(a), as we approach the slow synaptic limit ($\tau_m/\tau_s\to 0$) the resonant region (where the dominant eigenvalues are complex) requires increasingly larger values of $\eta$ and $\Delta$, vanishing for small enough ratio $\tau_m/\tau_s$.
Hence, as expected from the time scale analysis of Section~\ref{section:slowfast},
the dynamics of NMM2 can be faithfully reproduced by NMM1 in this limit.
However, the equivalence cannot be extrapolated
to physiological parameter values.

\subsection{Interneurons}\label{section:PV}

Here we consider a population of GABAergic interneurons with self-recurrent inhibitory coupling ($J<0$).
In particular, we focus on parvalbumin-postive (PV+) fast spiking neurons,
which play a major role in the generation of fast collective brain oscillations~\citep{bartos2002, bartos2007,cardin2009,tiesinga2009}.
We thus set $\tau_m=7.5$ and $\tau_s=2$ ms, following 
Table~\ref{tab:taus}.

In this case, the NMM1 dynamics are rather simple: there is a single fixed point that remains stable, with
a pair of complex conjugate eigenvalues (see Eq~\eqref{eq:eigvals}). Therefore, the transient dynamics do display resonant behavior upon external perturbation.
Nonetheless, no self-sustained oscillations emerge.

\begin{figure*}[t]
  \centerline{\includegraphics[width=\textwidth]{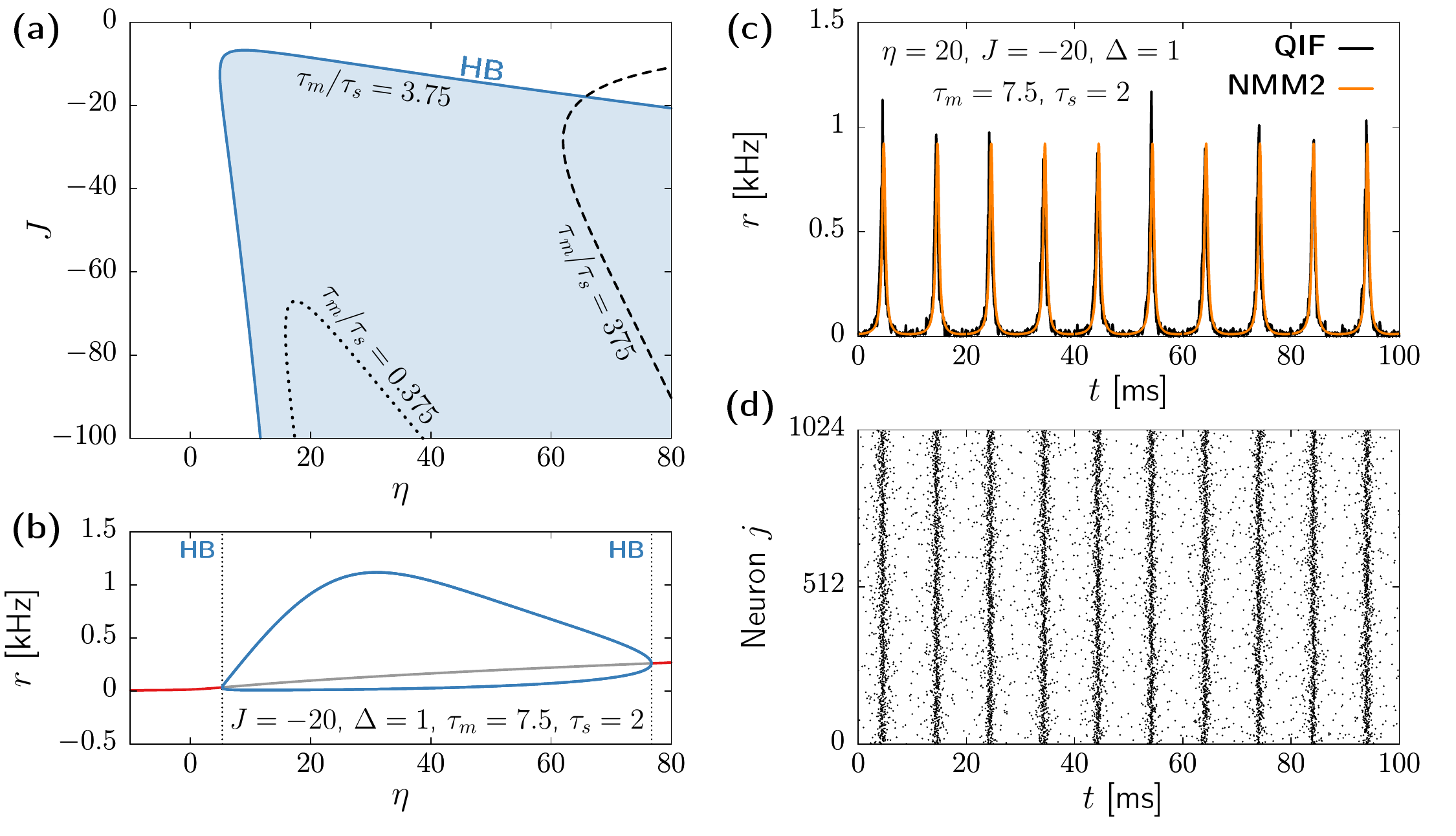}}
        \caption{
      {\bf Dynamics of a population of parvalbumin-positive interneurons described by NMM1 and NMM2}.
        (a) Supercritical Hopf bifurcation signaling the onset of oscillatory activity for $\tau_m=7.5$~ms, $\tau_s=2~ms$ (blue curve),
        $\tau_m=7.5$~ms, $\tau_s=20$~ms (dotted black curve), and $\tau_m=7.5$~ms, $\tau_s=0.02$~ms (dashed black curve).
        The blue-shaded region indicates stable limit-cycle behavior for $\tau_m=7.5$~ms, $\tau_s=2$~ms.
        (b) Steady-state values of the firing rate as a function of the input $\eta$, for fixed $J=-20$, $\Delta=1$, $\tau_m=7.5$~ms, and $\tau_s=2$~ms.
        The red line represents the stable steady state, the grey line the unstable steady state, and the blue lines the maxima and minima of the stable limit-cycle.
        Dashed vertical lines indicate the location of supercritical Hopf bifurcations (cf panel a).
        (c) Time evolution of the firing rate $r$ for an inhibitory population at the oscillatory state ($\eta=20$, $J=-20$, $\Delta=1$, $\tau_m=7.5$~ms, and $\tau_s=2$~ms)
        obtained from integrating the a network with $N=1024$ QIF neurons~\eqref{eq:qif} (black) and from the NMM2~\eqref{eq:nmm2} (orange).
        (d) Raster plot of the spiking times in the simulation of the QIF network corresponding to panel (c).
        Simulations of QIF network were performed with $V_\text{apex}=-V_\text{reset}=100$
        using Euler-Maruyama integration with $dt=10^{-3}$~ms.
        The firing rate $r$ was computed using Eq.~\eqref{eq:firingrate} with $\tau_r=10^{-2}$~ms.
        }
  \label{fig:PV}
\end{figure*}

In the NMM2, however, the unique fixed point might lose stability for $\eta>0$ through a
supercritical Hopf bifurcation (HB+, see blue curve in Fig.~\ref{fig:PV}(a)).
This transition gives rise to a large region of fast oscillatory activity,
corresponding to the so-called interneuron-gamma (ING) oscillations~\citep{whittington1995,traub1998,whittington2000,bartos2007,buzsaki2012}.
An example of this regime is shown in Figs.~\ref{fig:PV}(b,c), using both NMM2 as well as microscopic simulations of a QIF network as defined by Eq.~\eqref{eq:qif} .

According to the ING mechanism, oscillations emerge due to a phase lag between two opposite influences:
the noisy excitatory driving (controlled by $\eta$ and $\Delta$) and the strong inhibitory feedback from the recurrent connections (controlled by $J$).
In NMM2, the dephasing between these two forces stems from the 
implicit delay caused by the synaptic dynamics.
Hence, the ratio between membrane and synaptic characteristic times, $\tau_m$ and $\tau_s$, 
has a fundamental role in the generation of ING oscillations. The
blue region depicted in Fig.~\ref{fig:PV}(a) corresponds to time scales of PV+ neurons, $\tau_m=7.5$
and $\tau_s=2$. In this case the oscillation frequency is in the gamma range (40-200Hz). 
However, by decreasing the parameter ratio $\tau_m/\tau_s$ the Hopf bifurcation becomes elusive, as the oscillatory region shrinks, and oscillations require stronger inhibitory
feedback (see black dotted curve in Fig.~\ref{fig:PV}(a)).
Similarly, by increasing $\tau_m/\tau_s$ the ING activity also fades, as larger inputs $\eta$ are required to produce
oscillatory activity (see black dashed curve in Fig~\ref{fig:PV}(a)).
As showed in the previous section, the
two limits of $\tau_m/\tau_s$ coincide with NMM1 and
the model analyzed in~\cite{Montbrio:2015aa}.
The results presented above show that the membrane and synaptic dynamics are required to
have comparable time scales, in order to generate oscillatory activity in NMM2.

\subsection{Network-enhanced resonance in excitatory populations}\label{section:resonance}

The bifurcation analysis of Section~\ref{section:PN} reveals that a single population of excitatory neurons
does not display self-sustained oscillations in neither NMM1 nor  NMM2.
This is expected, as excitation alone is known to be usually insufficient for the emergence of collective rhythms~\citep{VanVreeswijk1994}.
However, in NMM2, the high-active steady state corresponds to a stable focus in a large region of the parameter space.
In this section we exploit this resonant behavior, inspired by the oscillatory response of a population of pyramidal neurons subject to tACS stimulation.
We thus consider the NMM2 model with $\tau_m=15$~ms and $\tau_s=10$~ms injected with a current
\begin{equation}\label{eq:tacs}
        I_E(t)=A\sin(\omega t)\;.
\end{equation}
We expect to induce oscillatory activity if $\omega$ is close to the resonant frequency of the system,
given by $\nu:=\Imag [\lambda]$, where $\lambda$ is the fixed point eigenvalue with largest real part.

\begin{figure*}[t!]
    \centering
    	\includegraphics[width=1\textwidth]{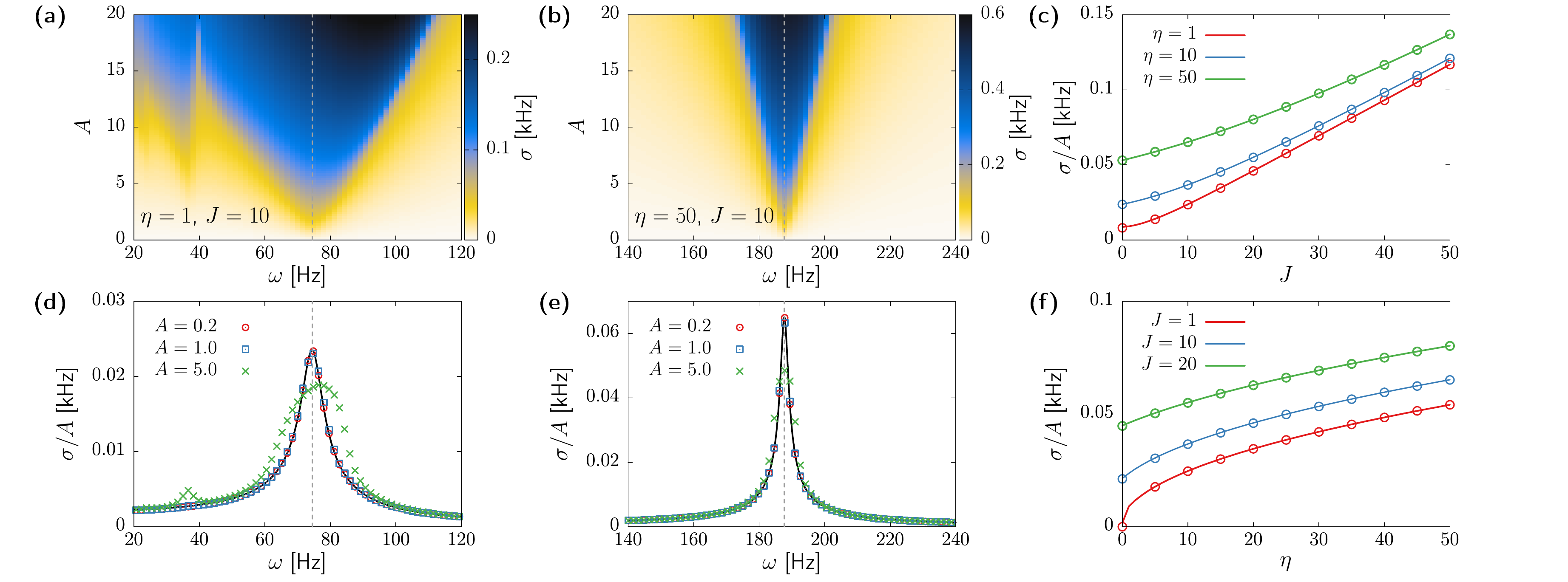}
    \caption{Effects of tACs stimulation, Eq.~\eqref{eq:tacs}, in a population pyramidal neurons given by NMM2~\eqref{eq:nmm2}.
    (a,b) Heatmaps of the standard deviation of $r$ displaying Arnold tongues for $\eta=1$ (panel (a))
    and $\eta=50$ (panel(b)). 
    The rest of system parameters are $J=10$, $\Delta=1$, $\tau_m=15$~ms, and $\tau_s=10$~ms. 
    (c) Normalized amplitude $\sigma/A$ obtained by stimulating the population at its resonant frequency $\nu$,
    for increasing values of the coupling strength. Continuous lines correspond to analytical results (Eq.~\eqref{eq:analytics}),
    and circles correspond to numerical simulations.
    (d,e) Normalized amplification $\sigma/A$ corresponding to the same parameters of panels (a) and (b), respectively.
    Symbols correspond to the numerical simulations reported in (a) and (b). The black continuous lines correspond to Eq.~\eqref{eq:analytics}.
    (f) Normalized amplitude $\sigma/A$ at the resonant frequency $\nu$ upon increasing the external input $\eta$.
    Lines correspond to Eq.~\eqref{eq:analytics} and symbols to numerical simulations.
    In all panels, periodic stimulation has been simulated for 2 seconds after letting the system relax to the fixed point for 1 second.
    The reported values for the standard deviation $\sigma$ correspond only to the last 1~s of stimulation, in order to avoid capturing transient effects.
    }
    \label{fig:resonance}
\end{figure*}

Figures~\ref{fig:resonance}(a,b) display heatmaps of the standard deviation of the firing rate, $\sigma$, obtained by stimulating the stable focus of NMM2 at different frequencies $\omega$ and amplitudes $A$.
For weak baseline input $\eta$ (Fig.~\ref{fig:resonance}(a)), the amplitude of the system displays a large tongue-shaped region, with a few additional narrow tongues at smaller frequencies.
The main tongue is centered at the resonant frequency $\omega\simeq\nu$ (see grey vertical dashed line) and corresponds to entrainment at the driving rhythm, whereas secondary tongues correspond to entrainment at higher harmonics.
Increasing the external input $\eta$ (Fig.~\ref{fig:resonance}(b)) causes the system to resonate at larger frequencies, and shrinks the region of amplification of the applied stimulus. 
Despite the similitude with the usual Arnold tongues that characterize driven oscillatory systems,
we recall that here we are inducing oscillatory activity in an otherwise stationary system.
Hence, even if small in amplitude, there is always
an oscillatory response at some harmonic of the driving frequency.

Electric stimulation protocols usually achieve large effects even when the amplitudes of the oscillatory input signal are small.
We thus investigate the effect of weak stimuli through a perturbative analysis for $0<A\ll 1$.
Upon expanding the NMM2 equations close to the fixed point and solving the resulting linear system,
we obtain the amplitude response 
as a function of the driving frequency:
\begin{equation}\label{eq:analytics}
\begin{aligned}
       \mathcal{A}&(\omega;\lambda,\beta) =\\
       & \Bigl\{   \left[ (\omega^2+\mu^2-\nu^2)(\nu b_\text{i}-\mu b_\text{r}) - 2\mu\nu(\nu b_\text{r}+\mu b_\text{i}) \right]^{2} \\
       &+ \omega^2\left[ 2 b_\text{i} \mu \nu + b_\text{r}(\omega^2+\mu^2-\nu^2) \right]^{2}\Bigr\}^{1/2}\\
       &\times  \left[ (\omega^2+\mu^2-\nu^2)^2+4\mu^2\nu^2\right]^{-1}
\end{aligned}
\end{equation}
where $\lambda=\mu+i \nu$ is the fixed point eigenvalue with largest real part,
and $b=b_r+ib_i$ is the associated amplitude component (see Appendix~\ref{secA2} for the mathematical details).
These two complex quantities can be obtained by numerically computing the eigenvalues
and eigenvectors of the system Jacobian.
The black curves in Fig~\ref{fig:resonance}(d) and (e) illustrate the validity of the analytical expression
when compared with numerical results (colored symbols).
Overall, the perturbative analysis provides a good approximation for $A<1$, showing that, at this stage, $\omega=\nu$
provides the maximal amplification.

Finally, we use these results to investigate the effect of the system parameters $J$ and $\eta$ 
to the amplitude response of the neural mass.
Figs.~\ref{fig:resonance}(b,c) show numerical (open circles) and analytical (lines) results obtained using the optimal stimulation
protocol $\omega=\nu$ with $A=0.1$ for different values of $J$ and $\eta$.
Overall, the oscillation amplitude of the system shows a supralinear increase with $J$, and a sublinear increase with $\eta$.
These results illustrate the importance of self-connectivity in tACs stimulation, and can potentially explain the effectiveness of these protocols in spite of the
weakness of the applied electric field.
Since we only considered driving of an excitatory population,
the associated resonant frequencies can be quite large (up to $400$Hz for $\eta=50$ and $J=50$),
which calls for future investigations to analyze the combined effect of tACs in networks
with excitation-inhibition balance.

%% file: conclusions.tex
\section{Conclusions}\label{section:conclusions}

For decades, NMMs have been built up on the basis of a simple framework that combines the linear dynamics of synaptic activation
with a nonlinear static transfer function linking neural activity (firing rate) to excitability~\citep{wilson_excitatory_1972,Freeman:1975aa,lopes_da_silva_model_1974}. 
This view has been sustained by empirical observations and heuristic assumptions underlying neural activity.
Models based on this framework have been used to explain the mechanisms behind neural oscillations~\citep{lopes_da_silva_model_1974,freeman_simulation_1987,jansen_electroencephalogram_1995,wendling_epileptic_2002}, and, more recently, to create large-scale brain models to address the treatment 
of neurophatologies by means of electrical stimulation~\citep{kunze_transcranial_2016,sanchez-todo_personalization_2018,forrester2020}.

Further theoretical efforts have provided more sophisticated tools to model the dynamics of neural populations,
by deriving transfer functions from specific single-cell models~\citep{gerstner1995,brunel2008,ostojic2011,carlu2020}, add adaptation mechanisms~\citep{augustin2017}, or finite size-effects~\citep{benayoun2010,buice2010}. 
In this context, exact NMMs (also known as Next-Generation NMMs) pave a new road to directly relate single neuron dynamics
with mesoscopic activity~\citep{Montbrio:2015aa}.
Understanding how this novel framework relates to previous semi-empirical models should allow us to validate the range of applicability of classical NMMs.

Here we have studied a neural mass with second-order synapses, similar to the one studied in recent works~\citep{coombes2019,byrne2020,byrne2022}.
The model naturally links the dynamical firing rate dynamics derived by~\cite{Montbrio:2015aa}
with the typical linear filtering representing synaptic transmission that is used in heuristic NMMs.
Following~\cite{ermentrout1994} and~\cite{Devalle:2017aa} we show that, in the slow-synapse limit and in the absence of time-varying inputs, the exact model can be formally mapped to a simpler formalism with a static transfer function. 
However, we find that the range of validity of this relationship is beyond the physiological values of the model parameters. 
An analysis of the dynamics using realistic values of the time constants illustrates the fact that 
fundamental properties, such as the resonant behavior of excitatory populations and the interneuron-gamma
oscillatory dynamics of PV+ neurons, cannot be captured by a traditional formulation of the model.

Despite the exact mean-field theory leading to NMM2 is a major step forward on the development of realistic mesoscale models for neural activity, the QIF neuron is a simplified model with some limitations. 
For instance, here we have employed non-refractory neurons,
for which increasing input currents always lead to an increase of the firing rate.
Future studies should address the role of a refractive period on the emerging rhythms and stimulation
effects of exact NMMs. This could lead to a more realistic saturating shape of the QIF transfer function (Fig.~\ref{figure0}).
Additionally,
further considerations may need to be taken into account in order to translate experimental observations to the model.
In particular, the synapse time constants reported in Table~\ref{tab:taus} should reflect the delay and filtering associated with current transmission from input site to soma. 
This is not trivial to measure experimentally, and it can change considerably depending on synapse location, morphology, the number of simultaneously activated spine synapses~\citep{Eyal2018}, and electrical properties~\citep{Koch2003-bi}, which are not accounted by the QIF neuron, but can be estimated using realistic compartment models~\citep{agmonsnir:1993aa}. Besides, the QIF model is an approximation of type-I excitable neurons, with type-II having a completely different firing pattern and $f$-$I$ curve. 

An important application of the exact mean-field theory is in the context of transcranial electrical stimulation.
 Several decades of research suggest that weak electric fields influence neural processing \citep{Ruffini:2020aa}.
 In tES, the electric field generated on the cortex is of the order of 1 V/m, which is known to produce a sub-mV membrane perturbation \citep{bikson_effects_2004,ruffini_transcranial_2013,aberra_biophysically_2018}. Yet, the applied field is mesoscopic in nature and is applied during long periods, with a spatial scale of several centimeters and temporal scales of thousands of seconds. Hence, a long-standing question in the field is how networks of neurons process spatially uniform weak inputs that barely affect a single neuron, but produce measurable effects in population recordings. 
 By means of the exact mean-field model, we have shown that the  sensitivity of the single population to such a weak alternating electric field can be modulated by the intrinsic self-connectivity and the external tonic input of the neural population in a population of excitatory neurons. 
 Importantly, such resonant behavior cannot be captured by heuristic NMMs with static transfer functions.
 
 For the physiologically-inspired parameter values chosen in this study,
the amplification effects on excitatory neurons appear to be weaker than those observed experimentally.
 We may conjecture that certain neuronal populations may be in  states near criticality, i.e. close to the 
 bifurcation points in the  NMM2 model \citep{RChialvo2004,Carhart2018,VzquezRodrguez2017,Zimmern2020-dz,Ruffini2022-wn}. This would apply, for example, to inhibitory populations, which display a Hopf bifurcation where a state near the critical point will display arbitrarily large amplified sensitivity to weak but uniform perturbations applied over long time scales. Since electric fields are expected to couple more strongly to excitatory cells, this case should be studied in the context of a multi-population NMM2, with excitatory cells relaying the electric field perturbation.
 Exact NMMs provide an appropriate tool to investigate this behavior, as well as the effects of non-homogeneous electrical fields---which we leave to future studies.

%% file: appendix.tex
\begin{appendices}

\section{Parameter reduction in NMM1 and NMM2}\label{secA1}

A common way to simplify the analysis of dynamical systems such as NMM1 (Eq.~\ref{eq:nmm1}) and 
NMM2 (Eqs.~\ref{eq:synaptic},\ref{eq:nmm2}) is through parameter reduction.
While this can be achieved in different ways, here we choose, following~\cite{Montbrio:2015aa},
to rescale the system parameters as follows:
\begin{equation}
\tilde \eta=\frac{\eta}{\Delta},\quad
\tilde J=\frac{J}{\sqrt{\Delta}},\quad
\beta=\frac{\tau_s\sqrt{\Delta}}{\tau_m}
\end{equation}
We then define the new variables
\begin{equation}
\begin{aligned}
\tilde t=\frac{\sqrt{\Delta}}{\tau_m}t,
&\quad\tilde r=\frac{\tau_m}{\sqrt{\Delta}}r,
\quad\tilde v=\frac{v}{\sqrt{\Delta}},\\
\quad \tilde s=\frac{ \tau_m }{\sqrt{\Delta}}s,
&\qquad \tilde z=\frac{ \tau_m }{\Delta}z\;.
\end{aligned}
\end{equation}
With these definitions, and together with the equivalence relation~\eqref{eq:equiv}, NMM1 takes the form:
\begin{equation}
\begin{aligned}
        \beta \frac{d\tilde s}{d\tilde t} &= \tilde z \\
        \beta \frac{d\tilde z}{d\tilde t}&=  \Psi_1[\tilde \eta + \tilde J \tilde s] - 2\tilde z -\tilde s\;.
\end{aligned}
\end{equation}
Similarly, the NMM2 model (Eqs.~\ref{eq:synaptic},\ref{eq:nmm2}) becomes
\begin{equation}
        \begin{aligned}\label{eq:reduced_fre}
                \frac{d\tilde r}{d\tilde t} &= \frac{1}{\pi} + 2\tilde r\tilde v\\
                \frac{d\tilde v}{d\tilde t} &= \tilde \eta - (\pi \tilde r)^2 + \tilde v^2 +\tilde J\tilde  s\\
          \beta  \frac{d\tilde s}{d\tilde t} &= \tilde z \\
          \beta  \frac{d\tilde z}{d\tilde t} &= \tilde r - 2\tilde z - \tilde s \;.
\end{aligned}
\end{equation}
These reduced systems reveal that the  dynamics of both models are controlled only by the three effective parameters $\tilde \eta$, $\tilde J$, and $\beta$. 
In particular, the effects of changing $\Delta$ in the attractors of the system can be achieved by appropriately 
modifying the other parameters. Also, this reduction makes explicit that the bifurcations
of the system do not depend on specific values of $\tau_m$ and $\tau_s$, but only on their ratio.
Notice, however, that in~\eqref{eq:reduced_fre} all parameters and variables, including time, become adimensional.
This contrasts with the formulation used throughout the paper (Eqs.~\eqref{eq:nmm2}), where time
has units of miliseconds.

\section{Analysis of a weakly periodically forced system}\label{secA2}

Here we present the results on weakly periodically perturbed systems used to investigate the response of NMM2 to periodic stimulation in section~\ref{section:resonance}.
Although we have a specific system in mind, we consider a general setup for simplicity.
Let $\bs{x}(t)\in\mathbb{R}^N$ be an $N$-dimensional state vector, with time evolution given
by the autonomous nonlinear system
\begin{equation}\label{eq:sys}
        \dot{\bs{x}} =\bs{F}(\bs{x})\;.
\end{equation}
In NMM2, $\bs{F}$ follows Eqs.~(\ref{eq:synaptic},\ref{eq:nmm2}), and the state vector reads $\bs{x}=(r,v,s,z)^T$.
Let $\bs{x}^{(0)}$ be a stable fixed point of the system and $\bs{J}=\bs{J}(\bs{x}^{(0)})$ the corresponding Jacobian.
We consider a periodic forcing acting on Eq.~\eqref{eq:sys},
\begin{equation*}\label{eq:per}
        \dot{\bs{x}} =\bs{F}(\bs{x})+\epsilon \bs{a}\sin(\omega t)
\end{equation*}
where $\epsilon$ is a weak coupling $0< \epsilon\ll 1$ and $\bs{a}\in \mathbb{R}^N$ is a normalized vector
for the distribution of the forcing across the system variables.
For instance, in the case considered in the paper $\bs{a}=(0,1,0,0)^T$, since the periodic driving
acts only on the mean membrane potential.

Let $\bs{x}=\bs{x}^{(0)}+\epsilon \bs{\delta x}$.
Since $\epsilon \ll 1$ we can linearize close to the fixed point, $\bs{x}^{(0)}$, to obtain
\begin{equation*}
        \dot{\bs{\delta x}}= \bs{J}\bs{\delta x} + \bs{a}\sin(\omega t)\;.
\end{equation*}
Let $\bs{S}$ be the matrix of eigenvectors of $\bs J$, and $\bs \Lambda$ the diagonal matrix
of eigenvalues, so that $\bs{S}^{-1}\bs{J}\bs{S}=\bs \Lambda$.
The coordinates of the perturbation vector in the basis defined by the Jacobian eigenvalues
read $\bs{\alpha}:=\bs{S}^{-1}\bs{\delta x}$. Therefore,
\begin{equation}
\begin{aligned}\label{eq:matrix}
        \dot{\bs{\alpha}}&=\bs{S}^{-1}\dot{\bs{\delta x}}\\
       & =\bs{S}^{-1} \bs{J}\bs{S}\bs{S}^{-1}\bs{\delta x} +  \bs{S}^{-1}\bs{a}\sin(\omega t)\\
        &= \bs{\Lambda} \bs{\alpha} +  \bs{b}\sin(\omega t)
\end{aligned}
\end{equation}
where $\bs{b}:=\bs{S}^{-1}\bs{a}$, i.e., the coordinates of $\bs{a}$ in the basis defined by $\bs{S}$.

Since $\bs{\Lambda}$ is a diagonal matrix, Eq.~\eqref{eq:matrix} can be written in scalar form for each $\alpha_j$ in complex space as
\begin{equation*}
        \dot \alpha_j = \lambda_j \alpha_j + b_j \sin(\omega t)\;, \quad\text{for}\quad j=1,\dots,N\;.
\end{equation*}
In what follows we drop the subindices $j$ for simplicity.
The solution of each of the linear systems for $\alpha$ read
\begin{equation*}
        \alpha(t)= -b \frac{\omega \sin(\omega t)+\lambda \cos(\omega t)}{\lambda^2 + \omega^2} + k e^{\lambda t}\;.
\end{equation*}
with $k \in \mathbb R$ a free constant. Since the fixed point is stable, the last term vanishes in the long term.
Let $\lambda=\mu+i\nu$. The behavior of $\alpha$ greatly changes depending on whether $\nu$ is zero or not,
i.e. whether the fixed point is a stable node or a stable focus.
Let us start for the simple case, $\nu=0$.
Then, at $t\to \infty$,
\begin{equation*}
        \alpha(t)=b\frac{\sin(\omega t + \phi)}{\sqrt{\mu^2+\omega^2}}
\end{equation*}
where $\phi=\arctan(\mu/\omega)$.
Therefore, these type of components always oscillate, but the amplitude of the oscillations decays as $1/\sqrt{\mu^2 + \omega^2}$.
Hence, if the forcing frequency is too fast, or the stability too strong, then the induced oscillatory component becomes negligible.\\

Let's turn now to the more interesting case of $\nu\neq 0$. Since we
are considering a real system, there is always a pair of complex eigenvalues such that $\lambda_\pm=\mu \pm i\nu$
associated to complex conjugate eigenvectors $\alpha_\pm$ (conjugate root theorem for polynomials with real coefficients).
Therefore the dynamics of the real system is given by the real part of $\alpha_\pm$.
We find that (for $t\to\infty$),
\begin{align*}
        \frac{\alpha(t)+\alpha^*(t)}{2}
        =\epsilon \mathcal{A}(\omega;\mu,\nu)\sin(\omega t+\phi)
\end{align*}
where
\begin{equation}\label{eq:amplitude}
\mathcal{A}(\omega;\mu,\nu)=\frac{\sqrt{M^2+N^2}}{D}
\end{equation}
and $\phi=\arctan(N/M)$, 
with
\begin{equation*}
\begin{aligned}
M&:=(\omega^2+\mu^2-\nu^2)(\nu b_\text{i}-\mu b_\text{r}) - 2\mu\nu(\nu b_\text{r}+\mu b_\text{i})\;, \\
N&:= \omega[- 2 b_\text{i} \mu \nu  -b_\text{r}(\omega^2+\mu^2-\nu^2)]\;,\\
D&:=(\omega^2+\mu^2-\nu^2)^2+4\mu^2\nu^2\;,
\end{aligned}
\end{equation*}
which corresponds to Eq.~\eqref{eq:analytics} in the main text.


\end{appendices}